\documentclass{mn2e}

\newif\ifAMStwofonts

\newcommand{\HI}{\mbox{H\,{\sc i}}}

\newcommand{\Msun}{M$_\odot$}

\title{Improving the triaxial bulge model of M31}
\author[Simon Berman and Laurent Loinard]
       {Simon Berman$^1$ and Laurent Loinard$^2$\\
        $^1$ Theoretical Physics, University of Oxford, 1 Keble Road, 
             Oxford, UK; simon@thphys.ox.ac.uk\\
        $^2$ Instituto de Astronom\'{\i}a, Universidad Nacional Aut\'onoma de 
             M\'exico, Apartado Postal 72--3, 58089 Morelia, Michoac\'an, 
             M\'exico}

\date{Accepted  .
      Received  ;
      in original form }

\pubyear{}

\usepackage{epsfig}

\begin{document}

\maketitle

\begin{abstract}
A detailed hydrodynamical model of the gas flow in the triaxial
gravitational potential of the bulge of the Andromeda galaxy (M31) has
recently been proposed by Berman (2001), and shown to provide
excellent agreement with the CO emission line velocities observed
along its major axis. In the present paper, we confirm the validity of
that model by showing that it can also reproduce the CO velocities
observed off the major axis -- a much more robust test. The CO
observations, however, tend to span a wider range of velocities than a
direct application of the original model of Berman would suggest. This
situation can be improved significantly if the molecular disk is made
thicker, a requirement already encountered in dynamical simulations of
other spiral galaxies, and typically attributed to a broadening of the
molecular layer in galactic fountain--like processes. In the central
regions of M31, however, it is unclear whether there actually is a
thick molecular disk, or whether broadening the molecular layer is
merely an artificial theoretical means of accounting for some disk
warping. Other effects not included in the model, such as hydraulic
jumps, might also contribute to a widening of the velocities.
\end{abstract}


\begin{figure*}
\centering
\epsfig{file=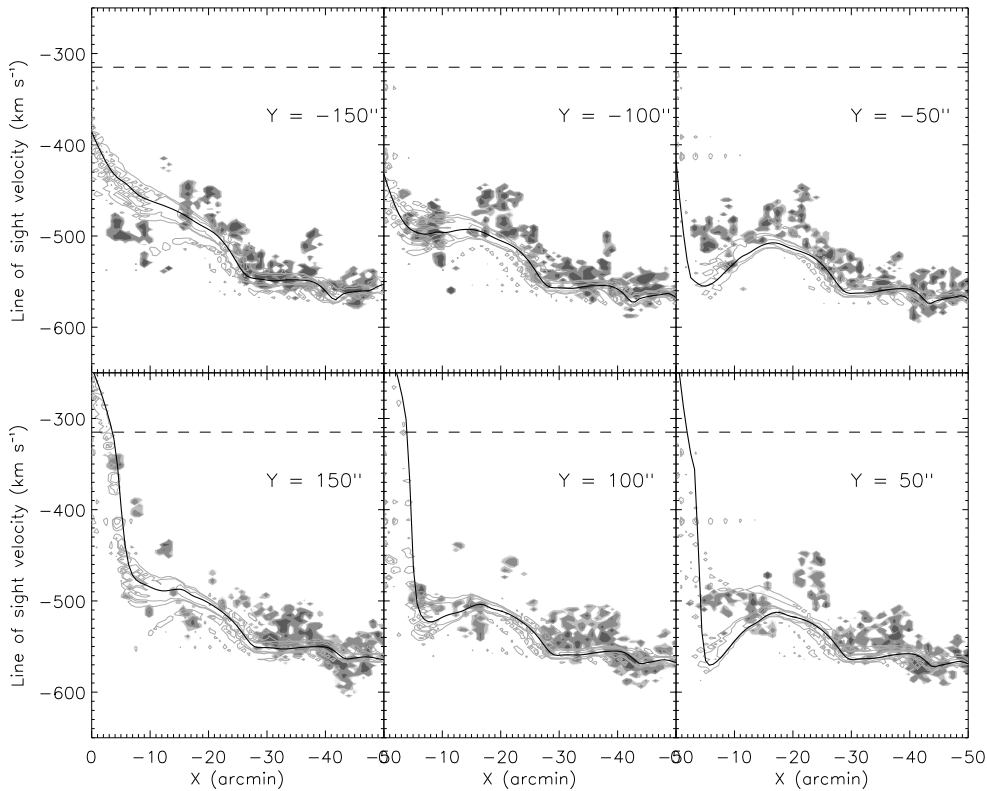, height=0.45\textheight}
\caption[Off axis observations]{Comparison of CO observations and gas
velocities from the model of maximum likelihood of Berman (2001) away
from the line of nodes of the disk. Model data were convolved with the
2D function representing the FCRAO beam as described in the
text. Intensity is plotted as a greyscale for observations and using
contour levels for model data. Contours are given at 1, 3, 10 and 30\%
of the maximum value in each subplot. The dashed line indicates the
systemic velocity and the solid line the results at $Y$ arcsec
ignoring the effects of the thick disk and of the beam function.}
\label{fig-offaxis}
\end{figure*}

\begin{figure*}
\centering
\epsfig{file=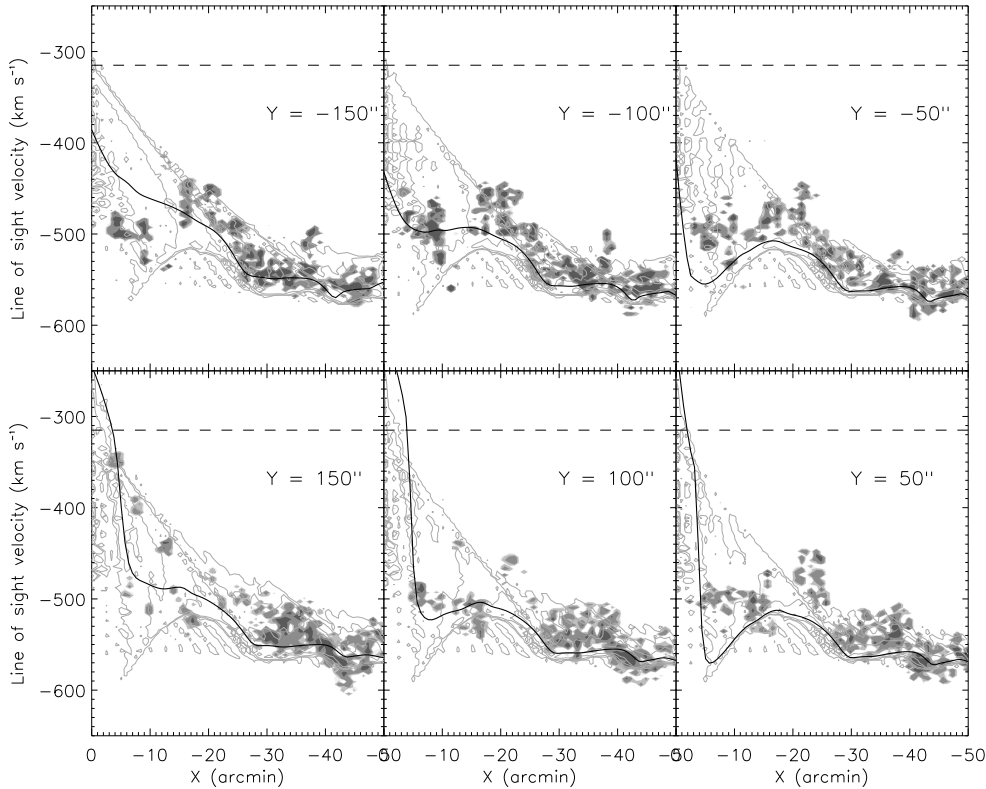, height=0.45\textheight}
\caption[Fast 3D model]{Comparison of CO observations and gas
velocities from the model of maximum likelihood away from the line of
nodes of the disk. The model includes a corotating thick disk of
scale height $z_0~=~400$ pc. Model data were convolved with the 2D
function representing the FCRAO beam as described in the
text. Intensity is plotted as a greyscale for observations and using
contour levels for model data. Contours are given at 1, 3, 10 and 30\%
of the maximum value in each subplot. The dashed line indicates the
systemic velocity and the solid line the results at $Y$ arcsec
ignoring the effects of the thick disk and of the beam function.}
\label{fig-fast3d}
\end{figure*}

\begin{figure*}
\centering
\epsfig{file=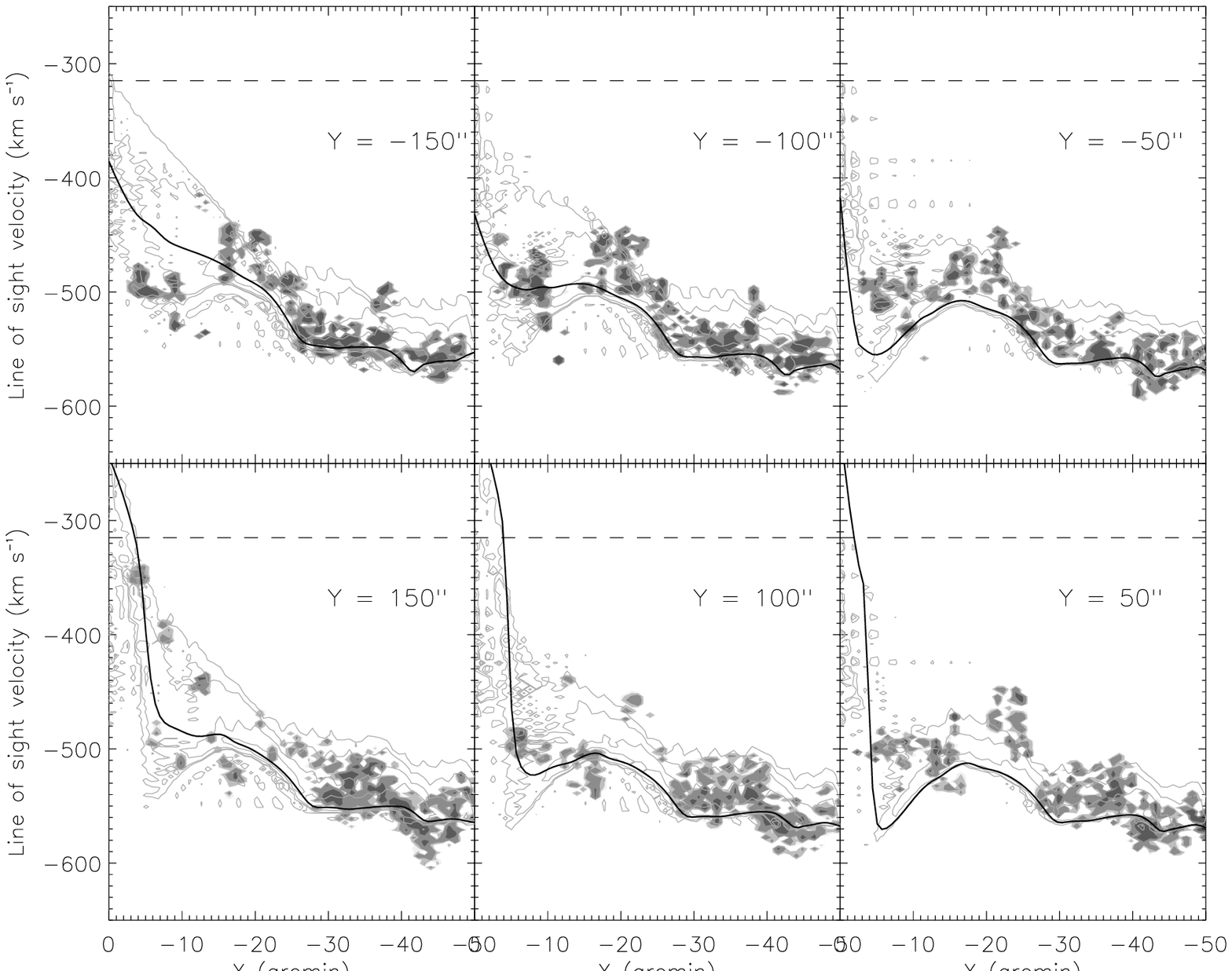, height=0.45\textheight}
\caption[Slow 3D model]{Comparison of CO observations and gas
velocities from the model of maximum likelihood away from the line of
nodes of the disk. The model includes a thick disk of scale height
$z_0~=~200$ pc and a loss of velocity of $0.3 \, {\bf v}_0$ kpc$^{-1}$
away from the plane of the disk. Model data were convolved with the 2D
function representing the FCRAO beam as described in the text.
Intensity is plotted as a greyscale for observations and using contour
levels for model data. Contours are given at 1, 3, 10 and 30\% of the
maximum value in each subplot. The dashed line indicates the systemic
velocity and the solid line the results at $Y$ arcsec ignoring the
effects of the thick disk and of the beam function.}
\label{fig-slow3d}
\end{figure*}

\section{Introduction}

There is growing evidence that spiral galaxies in which no bar is
easily seen on optical or infrared images, nonetheless, have triaxial
bulges. Twisting of the inner isophotes and misalignments between the
disk and the bulge major axes (both of which have been detected in
several spirals with no obvious bar) are clear signposts of
triaxiality. Even stronger cases have been made by combining optical
or infrared photometric data with spectroscopic observations of the
gaseous interstellar component. Indeed, when submitted to a triaxial
gravitational potential, the interstellar gas near the centre of a
spiral galaxy can be found at higher velocities than expected from
circular motion. Such anomalous velocities have been reported in
several galaxies where no strong bar is visible, including the two
nearest examples: the Milky Way (Rougoor \& Oort 1959; Dame et al.\
2001) and the Andromeda galaxy (Lindblad 1956; Loinard et al.\ 1999).

The triaxiality of spiral galaxy bulges is particularly important for
two reasons. First, it can affect the dynamics of interstellar gas,
gathering large amounts of it near galactic centres. This could help
to explain the appearance of active galactic nuclei. On long
timescales, a build of gas in galactic centres could affect the
morphology of spiral galaxies, and imply secular changes in their
Hubble types. Second, triaxiality allows more constraints to be
derived from spectral observations. In an axisymmetric system, the
rotation curve only allows the determination of the radial mass
distribution, without providing any three dimensional information. In
particular, it is not possible to assess in an axisymmetric system
whether the dark matter component is distributed in a flat disk-like
structure or in a spherical halo. In a triaxial system, that
disk--halo degeneracy can be removed, with obviously important
consequences for our understanding of dark matter. Cold Dark Matter
(CDM) models, for instance, favor spherical halos.

Since triaxiality primarily affects the dynamics at the centres of
spiral galaxies, it should be sought in our neighbours, where the
achievable linear resolution is highest. This is particularly
important when using radio observations of the gaseous component,
which often have limited angular resolutions. Detailed studies of the
dynamics of the central regions of the Milky Way are adversely
affected by our location inside of the system, and the well-known
associated confusion and distance ambiguity problems. However, the
existence of triaxial features in the inner Galaxy have been
postulated for over forty years. Based on the discovery of the 3 kpc
arm, Kerr (1967) proposed the existence of a gaseous bar at the Galaxy
centre, and Mulder \& Liem (1986) modelled the arm as a density wave
in a barred potential. By combining photometric and kinematic data,
Gerhard \& Vietri (1986) demonstrated the triaxiality of the Galactic
bulge, and COBE images of the Galactic centre have confirmed this
triaxiality (Dwek et al.\ 1995; Binney et al.\ 1997).

The next nearest spiral (the Andromeda Galaxy -- M31) has no clear
visible bar, and is an excellent target for a dynamics study, because
its high inclination enables an accurate determination of the gas
kinematics. M31 has long been known to have twisted inner isophotes
and misaligned bulge and disk major axes (Lindblad 1956). Moreover,
the existence of anomalous gas velocities in the inner few kiloparsecs
is also now well established both for the atomic (Brinks \& Burton
1984) and molecular (Loinard et al.\ 1996, 1999) components. A
convincing interpretation in terms of triaxiality was put forward as
early as 1956 by Lindblad, and was further developed by Stark (1977)
and Stark \& Binney (1994). However, those early works were hampered
by the lack of strong observational constraints.

The distribution and kinematics of the atomic component of the
interstellar medium (ISM) has long been known at high resolution and
sensitivity across the entire disk of M31 thanks to interferometric
observations of the 21--cm line of atomic hydrogen (Brinks \& Shane
1984, and references therein). However, those observations do not
provide clear information on the ISM kinematics in central regions,
because the kinematic component associated with the warped outer disk
can be seen in projection through the inner disk. The resulting
confusion between inner and outer disk seen in the \HI\
position-velocity diagrams of the central regions of M31 precludes
accurate studies of its dynamics. The molecular component traced by CO
emission is a better tracer of the dynamics of the inner regions
because no CO emission can be detected in the warped outer
disk. However, the inner regions of M31 are also particularly dim in
CO (Dame et al.\ 1993, Loinard et al.\ 1999). Fragmentary CO
observations of the central regions of M31 were used to constrain the
model of Stark \& Binney (1994). Recently, however, more systematic CO
observations have been obtained.

Loinard et al.\ (1995) presented a deep search for CO emission along
the inner major axis of M31. This search confirmed the existence of
anomalous velocities in the inner few kiloparsecs of M31, and results
along the line of nodes of the disk were used by Berman (2001) to
constrain an improved dynamical model of the central regions of M31
(see \S 2 below). However, Berman (2001) pointed out that off-axis CO
observations would provide better constraints. Such data are now
available -- at least for the southern half of M31 -- thanks to the
CO(1-0) survey made at the {\em Five College Radio Astronomy
Observatory} (FCRAO -- Loinard et al.\ 1996, 1999). The angular
resolution of the survey is $\sim$ 1 arcmin, and it covers the entire
Southern part of M31 with a sampling of 50 arcsec. The observational
noise level is rather constant across the entire surveyed region with
a typical r.m.s. of 45--50 mK per 3.25 km s$^{-1}$ spectral
channel. Although the CO(1-0) is found to peak in the broad Population
I ring at 10 kpc from the centre, significant emission is still
detected in the inner 2--3 kpc (10--15 arcmin), where most constraints
can be obtained about the triaxiality of the bulge. The integrated
intensity images or position--velocity diagrams shown in Loinard et
al.\ (1999), as well as the position--velocity diagrams that will be
shown in this article, have been obtained after `unsharp masking' was
applied to the data. In this scheme, a data cube smoothed both in
position and velocity (but not regridded) is first constructed from
the original data. The average noise level $\sigma$ in this smoothed
cube is computed, and all the pixels in the non--smoothed data cube
where no significant ($>$ 3$\sigma$) emission is found in the smoothed
version are blanked. This avoids that unnecessary noise is added when
summations over several pixels are performed, and in effect reduces
the noise in integrated maps by a factor of a few. A slight drawback
of this method is that the noise becomes dependent on the number of
pixels summed (or averaged), and, therefore, on the position in the
maps.

Anomalous velocities are clearly seen in this data set in the inner
regions of M31 (see fig.\ 12 in Loinard et al.\ 1999). While the cut
along the major axis essentially shows the same anomalous velocities
as reported by Loinard et al.\ (1995), cuts parallel to the major axis
provide new constraints for the models, as called for by Berman
(2001).

\section{Original model}

The dynamical model developed by Berman (2001) takes into account a
rotating, triaxial bulge, and an axisymmetric component that mimics
the combined influence of the disk and halo. The properties of the
bulge are obtained by finding a triaxial mass distribution which gives
the observed $R^{1/4}$ $B$-band surface brightness profile when
rotated and inclined to the plane of the sky. Properties of the
axisymmetric component are deduced from a simple interpretation of
\HI\ and CO kinematics. The gas response to the gravitational
potential is computed using a hydrodynamical code ({\sc galahad}) that
solves the non-self gravitating, isothermal Euler equations, on a
regular Cartesian grid, using the FS2 algorithm of van Albada et al.\
(1982). Each grid cell is 125 pc (37.5 arcsec) on a side, which is
similar to the resolution ($\sim$ 60 arcsec) of the CO data. Removal
and injection of interstellar matter by star--formation and stellar
mass-loss respectively are included (somewhat crudely) in the
algorithm.

Berman (2001) constrained the parameters of his model by comparing the
output gas flow to the CO observations along the line of nodes of the
disk of M31 of Loinard et al.\ (1995). The best fit to the data is
obtained for a bulge semi-major axis $a = $ 3.5 kpc, a pattern speed
$\Omega_p = $ 53.7 km s$^{-1}$ kpc$^{-1}$, a corotation radius to
bulge semi-major axis ratio $\mathcal R =$ 1.2, an angle $\phi$
between the major axis of the bulge and the disk's line of nodes of
15$^{\circ}$, and a bulge mass-to-light ratio $\Upsilon_B$ = 6.5.
This in turns implies a bulge mass\footnote{The value given for the
mass of the bulge in Berman (2001) was not correct. The amended
version has been used in this paper.} of $M_{\rm bulge}$ =
2.3~$\times$~10$^{10}$ \Msun. The disk mass within the radius of the
bulge (3.5 kpc) can be estimated by assuming reasonable values for the
$B$-band exponential disk central surface brightness ($I_0 = 21.6$ mag
arcsec$^{-1}$, Walterbos \& Kennicutt 1988) which when corrected for
absorption and inclination using the values of Berman (2001) gives a
central surface density of $\Sigma_d = $113 L$_{\odot}$ pc$^{-3}$,
disk scale length ($R_d = 5.8$~kpc, Walterbos \& Kennicutt 1988), and
a conservative disk mass-to-light ratio of 4. This gives a disk mass
of $M_{\rm disk}$ = 1.2~$\times$~10$^{10}$ \Msun. The total (bulge +
disk + halo) mass inside the same radius can be estimated from the
rotation curve as $M_{\rm total}$ = 3.7~$\times$10$^{10}$ \Msun,
leaving little room for a dark halo.

The value of $\mathcal R = 1.2$ found by Berman (2001) implies that
the bulge of M31 is a fast rotator. Interestingly, of the few galaxies
for which $\mathcal R$ has been measured, all have been found to be
fast rotators. Debattista \& Sellwood (2000) showed that such fast
bars cannot co-exist with massive dark halos because dynamical
friction would slow them down rapidly. The value of $\mathcal R = 1.2$
is further evidence that M31 must have a minimal halo, and,
consequently, a maximum disk. This is notably at odds with the
contentions of Cold Dark Matter (CDM) models, which predict massive
halos and minimal disks (e.g.\ Navarro et al.\ 1997).

\section{Modelling off--axis observations}

As mentioned earlier, Berman (2001) pointed out that a much stronger
case could be made for his best fit model if it were found to be
consistent with observations of CO velocities away from the line of
nodes of the disk.  Moreover, it is important to utilize these
off--axis velocities since dynamical models of the bulge of M31 have
never before been constrained by real 3D $(X,Y,V)$ observations.

To compare accurately the output of the model with the data, we
performed a 2D convolution of the model with a realistic description
of the FCRAO beam, consisting of the sum of a sinc function and two
Gaussians. The sidelobes of the sinc function provide a good (very
conservative) description of any coma effects, while the two Gaussians
describe the response of the FCRAO antenna to emission coming from
angles far away from the nominal pointing position, due to the
imperfections of the dish (Ladd \& Heyer 1996). The sinc function has
a full width at half maximum (FWHM) of 45 arcsec and the two Gaussians
have FWHM and attenuations of 30 arcmin and 30 dB, and 4 degrees and
50 dB, respectively. This convolution produces multiple velocities at
each $X$ position in the $(X,V)$ position--velocity diagrams derived
from the model, in the same way as multiple velocities arise naturally
in the data cube.

Fig.\ref{fig-offaxis} shows position--velocity diagrams for
observational data at $Y = \pm \, y \times 50$ arcsec, where $y = 1,
2, 3$. Overlayed are Berman's best fit model's predictions convolved
with the FCRAO beam function which should, if the model is correct,
approximately enclose the observations on every plot. In this, as in
all of the comparisons below, the model has been interpolated to fit
the positions of the observations. Although the general trends of the
predictions and the observations are similar, the model data enclose a
narrower range of velocities than the observations and are further
from the systemic velocity than the observations. The discrepancies
are just as large or greater for models with different input
parameters. In particular, it is not reasonable to use a smaller
rotation velocity. This would help reconcile the predictions of the
model proposed by Berman (2001) with the data in the central regions,
but would predict velocities much too small at distances larger than
about 40 arcmin from the centre.

The same `beard' phenomenon is seen in the position--velocity diagrams
of the spiral galaxy NGC~2403 (fig.\ 1 of Schaap et al.\ 2000). The
arguments presented in that paper conclude that the beard is either
caused by a thick high density gas layer with a half width at half
maximum (HWHM) intensity of 500 pc or a lower density layer with HWHM
of 1.75 kpc, rotating more slowly than the standard thin disk. This
conclusion is motivated by galactic fountain models (Spitzer 1990) in
which hot gas from supernova explosions and galactic winds rises from
the disk into the halo. As it rises, the gravitational attraction
towards the centre of the galaxy lessens, the gas moves outwards and,
by conservation of angular momentum, its azimuthal velocity
decreases. Thick molecular disks have been seen in CO in the Milky Way
Galaxy (Dame \& Thaddeus 1994), and the edge--on spiral galaxy NGC~891
(Garcia-Burillo et al.\ 1992).

To recover the small magnitude and large range of velocities of the
observations of M31, the bulge model of Berman (2001) is augmented by
the kinematics of a thick disk which has been rotated and inclined to
the observer's frame. The vertical density distribution follows Schaap
et al.\ (2000) in taking a Gaussian form
\begin{eqnarray}
\rho(z) & = & \rho_0 \, \exp(-z^2/2 \, z_0^2), \\
\rho_0  & = & \frac{\Sigma_0}{(2 \pi)^{1/2} \, z_0},
\end{eqnarray}

\noindent
where $\rho_0$ and $\Sigma_0$ are the density and surface density
respectively at $z = 0$, and $z_0$ is the vertical scale height of the
gas. The half width at half maximum intensity $z_{1/2} = 1.18 \,
z_0$. The intensities at $(X,Y)$ are convolved in 2D with the function
representing the FCRAO beam as described earlier, to produce
position--velocity plots for $Y = \pm \, y \times 50$ arcsec, where $y
= 1,2,3$. Fig.\ \ref{fig-fast3d} compares the observational data with
the results from a model with $z_0~=~400$ pc, the lowest value of the
scale height that adequately encloses the vast majority of the
observations for $X > -50$ arcmin. For this model, $z_{1/2}$ = 472
pc. This is very large when compared to the thick disk observed by
Dame \& Thaddeus (1994) in the Milky Way, for which HWHM is between 71
and 133 pc depending on position, but smaller than the thick disk of
NGC~891 (Garcia-Burillo et al.\ 1992), where HWHM is larger than 1
kpc.

We now follow Schaap (2000) and drop the implicit assumption that the
thick disk is corotating with the $z~=~0$ plane and reduce the
velocities of the gas where $z~\neq~0$. Velocities parallel to the $z$
axis stay fixed at zero. A slow thick disk is implied by the galactic
fountain models and was also invoked by Swaters et al.\ (1997) to
explain the \HI\ observations of NGC~891. We take the linear velocity
distribution used in both papers:
\begin{eqnarray}
{\bf v}(z) & = & {\bf v}_0 (1 - \gamma \, z),
\end{eqnarray}

\noindent
where ${\bf v}_0$ is the velocity in the $z = 0$ plane, $\gamma$ is
the velocity lost per kpc and $z$ is the height. By slowing the gas
down above and below the plane of the disk, the range of velocities at
a particular scale height is extended. It can be seen in Fig.\
\ref{fig-slow3d} that a similar range of velocities to both the
observations and the corotating model is produced if the scale height
is just 200 pc and $\gamma = 0.3$ i.e. the gas loses 30\% of its speed
per kpc away from the plane of the disk. This implies that
$z_{1/2}$~=~236 pc which is closer to but still larger than the values
quoted for the Milky Way in Dame \& Thaddeus (1994) and much smaller
than the value for NGC~891.

By including a thick disk, the model initially proposed by Berman
(2001) is able to match the 3D CO data of Loinard et al.\ (1999)
reasonably well. The most noticeable discrepancy is found around $X =
\pm 20-25$ arcmin, $V \sim~-$45~ km s$^{-1}$, a position that the
original model found difficult to account for even along the line of
nodes of the disk. It is plausible that some vertical or turbulent gas
motions not included in our model can account for these remaining
discrepancies.

As mentioned above, galactic fountain models are often invoked to account 
for thick gaseous disks in the centres of galaxies. However, they are 
probably not justifiable in M31 since it does not harbour a great deal of 
star formation. Moreover, whilst the thick disk found by Dame \& Thaddeus
(1994) in the Milky Way appears to be much fainter in CO than the
Galactic mid--plane, the present modelling requires that in M31 it be
just as bright. A plausible alternative follows from the recent work by 
Martos \& Cox (1998), who describe many numerical simulations combining 
magnetic fields with hydrodynamics to create postshock regions in which 
large pressure gradients exist near to the mid--plane causing the gas layer 
to expand vertically. The result is a `hydraulic jump' in which a build up 
of gas in the vertical direction can be seen with scale heights of hundreds 
of parsecs.

Another, perhaps more controversial, possibility should be mentioned
here.  Pringle et al. (2001) contend that much of the ISM in spiral
galaxies is molecular, but too cold to be easily detectable in CO. In
this view, only the parts of molecular clouds illuminated by newly
formed stars are bright in CO. The CO--bright thin disk could then
correspond to the thin layer where stars form most actively, whereas
the thick disk would correspond to the real total extent of the
molecular layer. The existence of large quantities of cold molecular
gas in the inner regions of M31 has been proposed by Loinard \& Allen
(1998 -- and references therein). Moreover, since the inner region of
M31 harbours little star--forming activity, the thin disk there would
not be particularly brighter in CO than the thick disk, and the
distinction between the two would indeed vanish.

Finally, a last possibility is that the inner disk of M31 is actually thin, 
but warped. Modelling it as a thick disk then becomes a convenient way to 
account for the gas located in the warped parts of the disk, above or below 
the mid--plane.

\section{Conclusions}

In this article, we show that hydrodynamical triaxial models of the
bulge of the Andromeda galaxy (M31) previously presented by Berman
(2001), and initially tested against CO emission observations along
the apparent major axis only, can also account for the CO emission
velocities observed off the apparent major axis -- a much more robust
validity test. The triaxial model implies that the bulge of M31 is a
fast rotator and hence the dark matter contained in its central
regions must be minimal and coincident with the stellar disk.

To account for the whole velocity range covered by the observations, a
finite, fairly large thickness (half width at half maximum intensity
of 200--500 pc) must be given to the molecular disk. It is quite
unclear, however, whether the inner molecular disk of M31 is truly
that thick or whether adding it to the model is merely an effective
theoretical trick able to account for a warping of a thiner
disk. Other out of plane, or peculiar velocities (such as those
associated with hydraulic jumps or turbulence) could also contribute
to the wide velocity range covered by the CO observations.

\section*{Acknowledgments}
We thank James Binney and Daniel Pfenniger for careful reading of the
manuscript, and interesting comments.

\end{document}